\documentclass[prl,aps,twocolumn,groupedaddress,floats,nofootinbib,,preprintnumbers]{revtex4}
\usepackage{graphicx}
\usepackage{float}
\usepackage{amsmath,amssymb,amsfonts}
\usepackage{bbm}

\newcommand{\bse}{\begin{subequations}}
\newcommand{\ese}{\end{subequations}}
\newcommand{\be}{\begin{equation}}
\newcommand{\ee}{\end{equation}}
\newcommand{\bea}{\begin{eqnarray}}
\newcommand{\eea}{\end{eqnarray}}
\newcommand{\ba}{\begin{array}}
\newcommand{\ea}{\end{array}}

\input amssym.def
\input amssym.tex

\usepackage[colorlinks=true, linkcolor=blue, bookmarks=true, citecolor=red]{hyperref}

\begin{document}

\title{Black Holes Evaporate Non-Thermally}

\author{Davood Allahbakhshi\footnote{allahbakhshi@ipm.ir}}
\affiliation{School of Particles and Accelerators, Institute for Research in Fundamental Sciences (IPM),
P.O.Box 19395-5531, Tehran, Iran}

\begin{abstract}
It is shown that the Hawking radiation and staying at the Schwarzschild state are not consistent for black holes as graviton balls. The only exception is the eternal black hole.
\end{abstract}


\maketitle

\section{Introduction}
In 1975 Hawking showed that eternal black holes radiate thermally \cite{Hawking:1974sw}. Such thermal evaporation makes the evolution from pure states to mixed states possible. A black hole made of a lump of matter in a pure state seems to evaporate completely to particles in thermal state, which is not possible by unitary evolutions. This is the famous information paradox \cite{Hawking:1976ra}.

The Hawking's calculation is done for an eternal black hole. Mass of the black hole does not change and the radiation is informationless. Many physicists believe that including the backreaction of the radiation on the black hole makes the radiation non-thermal and so not completely informationless \cite{Dvali:2015aja}. It may probably solve the information paradox or at least makes it less paradoxical.

A new picture of black holes is proposed \cite{Dvali-Gomez-papers} which considers them as graviton balls in a special state. It is proposed that the black hole is a ball of gravitons stuck at the critical point of a quantum phase transition.

This picture is recently improved and a little changed by considering black holes as \emph{gravitational Schwinger balls of gravitons} \cite{Allahbakhshi:2016ctk}. This is a region in space, full of gravitons, in which the gravitational field is stronger than the Schwinger limit of the gravitons which live in the region. It is shown that all characteristic equations of Schwarzschild black holes can be derived by using Newton's law coupled to Schwinger limit.

In present paper we push the picture a little forward by studying the evolution of these gravitational Schwinger balls. We will see that the Schwarzschild black hole and the thermal radiation with Hawking temperature are not consistent.

\section{Schwinger Ball of Gravitons}
As mentioned, recently it is shown that black holes can be considered as Schwinger balls of gravitons in the sense that they are regions in which the gravitational field can create gravitons which can live inside the region \cite{Allahbakhshi:2016ctk}.

We define the \emph{Schwarzschild state} with two relations below between the mass ($M$), size ($R$) and the number of gravitons ($N$) in the black hole
\bea
N &&= \frac{R^{d-2}}{L_P^{d-2}}\cr\cr\cr
R^{d-3} &&= \frac{GM}{c^2},
\eea
where $d$, $G$, $c$ and $L_P$ are the number of space-time dimensions, Newton's constant, speed of light and Planck length respectively. The name comes from the fact that the Schwarzschild black hole obeys these relations. In fact they determine the state of the Schwarzschild black hole.

Now let us consider a black hole of size $R$, with mass $M$ made of $N$ gravitons. Suppose that two gravitons inside the black hole decay to two couples of softer gravitons. Two of them (Bobs) escape from the black hole and other ones (Graces) will be trapped inside it. Since the wavelengths of Graces are larger than the size of the black hole, they have to escape from the black hole or merge. Suppose that they merge to produce a new Grace. This merging event reduces the entanglement entropy of the black hole by one unit \cite{Allahbakhshi:2016ctk}.

The very important fact is that \emph{if the black hole, after radiation, wants to stay at the Schwarzschild state then the total energy of Bob gravitons can not be an arbitrary value.} For calculating this allowed value of energy, suppose that the total mass of the black hole before and after emitting the gravitons are $M_1$ and $M_2$ respectively. So the total energy of the Bob gravitons is
\be
w = \big(M_1-M_2\big)c^2.
\ee
By using the relation $R^{d-3} = GM/c^2$, we have
\be
w=\frac{c^4}{G}\big[ R_1^{d-3}-R_2^{d-3}\big].
\ee
The energy $w$ in terms of the number of gravitons is
\be
w=\frac{c^4L_P^{d-3}}{G}\big[ N_1^{\frac{d-3}{d-2}}-N_2^{\frac{d-3}{d-2}}\big].
\ee
But $N_2=N_1-1$, so we have
\bea\label{non-thermal-main}
w = M_Pc^2 N^{\frac{d-3}{d-2}}\left[1-\left( 1-\frac{1}{N} \right)^{\frac{d-3}{d-2}}\right],
\eea
which can be rewritten as
\be\label{non-thermal-main-prime}
w= \frac{\hbar c}{R} \; f(N),
\ee
where
\be
f(N)=N\left[1-\left( 1-\frac{1}{N} \right)^{\frac{d-3}{d-2}}\right].
\ee
We have renamed $N_1\rightarrow N$. If the black hole wants to stay near the Schwarzschild state, up to fluctuations, then this energy must be the energy of \emph{the most probable radiated quanta}. In fact the black hole stays near the Schwarzschild state if and only if the energy of the radiated quanta obeys the relation above. If not, the black hole will be pushed away from the Schwarzschild state as time goes.

The relation \ref{non-thermal-main-prime} is very important. Since $f(N)\neq 1$, it can be interpreted as a deviation from thermality. For this to be more clear let us consider a large macroscopic black hole in which $N \gg 1$. For such a large black hole we can expand the relation\footnote{We have dropped the numerical coefficients for simplicity.}
\bea\label{non-thermal-series}
w &&=\frac{\hbar c}{R}f(N)\cr\cr
&&=\frac{\hbar c}{R}\left[1+\frac{1}{N}+\mathcal{O}(N^{-2})\right]\cr\cr
&&=\frac{\hbar c}{R}\left[1+\frac{1}{S}+\mathcal{O}(S^{-2})\right],
\eea
where $S$ is the entropy of the black hole. This relation says that the radiation is thermal as Hawking calculated, if and only if $N\rightarrow \infty$ at which the black hole becomes \emph{eternal}. For a finite size black hole with finite number of gravitons, as black hole evaporates, the radiation goes away from thermality. As mentioned previousely, if we want to keep the radiation thermal, with Hawking temperature, then the energy \ref{non-thermal-main} will not be the energy of the most probable radiated gravitons and so the black hole goes away from the Schwarzschild state which may not be the solution of Einstein-Hilbert action anymore. Our result is comparable to the results of \cite{Dvali:2015aja,Chakraborty:2016fye}.

Another possible interpretation of the equation \ref{non-thermal-main-prime} is that the radiation is still thermal and $\hbar c f(N)/R$ is a \emph{modified Hawking temperature}. This interpretation is also possible since the calculation above does not say anything about the spectrum of the radiation. For ruling out this claim we need to calculate the spectrum of radiation in this picture. But usual belief is that the radiation of a non-eternal black hole is not thermal.

\section{Thermalization After Radiation}

The energy of the new Grace inside the black hole after radiation is
\be
E_G=\frac{\hbar c}{\lambda}=\frac{2c\hbar}{R_1}-w.
\ee
You can check that the wavelength of the new Grace is shorter than $R_2$. So just after radiation we have $N-2$ gravitons with wavelength $R_1$ and a graviton with wavelength $\lambda<R_2$. The new Grace will pass its energy to other gravitons by interacting with them. Its wavelength becomes longer and the wavelengths of other gravitons become shorter. After a while the system will be thermalized at radius $R_2$. It should be possible to estimate this thermalization time.

During this thermalization process the system will experience an out of equilibrium evolution. There should not be a well defined horizon because the energy and so wavelength of all gravitons are not equal.

After thermalization the graviton system will have a definite size and so entropy but, continuous radiation of quantas always keep it away from equilibrium. This deviation from equilibrium may be more important when the black hole becomes small, since the radiated quanta has larger energy and the rate of the radiation is probably increased, although the system that should become thermalized, is smaller.

\section{Summary and Discussion}

In this paper we showed that the Schwarzschild state and thermality of the Hawking radiation with Hawking temperature are generally not consistent except for eternal black hole. Hawking radiation pushes the black hole away from the Schwarzschild state. On the other hand, keeping the black hole at the Schwarzschild state makes the radiation non-thermal or at least modifies the Hawking temperature.

The simple calculation which led to equation \ref{non-thermal-main} has three results. At first it reproduces the Hawking temperature, as the energy of the most probable radiated quanta\footnote{Remember that from Wien's displacement law we know that for thermal radiation the energy of the most probable radiated quanta is the temperature of the source.}, at the limit of eternal black hole. Secondly the equation \ref{non-thermal-main} is the energy of the most probable radiated quanta for a non-eternal black hole which wants to stay at Schwarzschild state. Such calculation, if possible, is so hard by semi-classical approaches. The third result is inconsistency between Schwarzschild state and Hawking radiation.

As mentioned previousely, after radiation the black hole will experience a period of out of equilibrium evolution, before it can become thermalized again. It may be very interesting and important if we can estimate this thermalization time. On the other hand continuous radiation keep the black hole out of equilibrium and Schwarzschild state. It will be also interesting if we can study such non-equilibrated black holes.



\end{document}